\documentclass[12pt,english,prb,reprint,superscriptaddress,nofootinbib,twocolumn]{revtex4-2}

\usepackage[T1]{fontenc}
\usepackage[latin9]{inputenc}
\usepackage{xcolor}
\usepackage{pdfcolmk}
\usepackage{amsmath}
\usepackage{amssymb}
\usepackage{graphicx}
\PassOptionsToPackage{normalem}{ulem}
\usepackage{ulem}

\makeatletter

\providecolor{lyxadded}{rgb}{0,0,1}
\providecolor{lyxdeleted}{rgb}{1,0,0}

\DeclareRobustCommand{\lyxsout}[1]{\ifx\\#1\else\sout{#1}\fi}

\usepackage[final]{changes}
\usepackage{ulem}
\usepackage{bbm}

\makeatother

\usepackage{babel}
\begin{document}
\title{Refined Majorana phase diagram in topological insulator-superconductor
hybrid system}
\author{Xin Yue}
\affiliation{Beijing Computational Science Research Center, Beijing 100193, China}
\author{Guo-Jian Qiao}
\affiliation{Beijing Computational Science Research Center, Beijing 100193, China}
\author{C. P. Sun}
\email{suncp@gscaep.ac.cn}

\affiliation{Beijing Computational Science Research Center, Beijing 100193, China}
\affiliation{Graduate School of China Academy of Engineering Physics, Beijing 100193,
China}
\begin{abstract}
The edge state of the topological insulator coupled to a superconductor
system is able to simulate the Majorana fermion in zero energy mode
since the Kitaev-type pairing is induced by exchanging quasi-excitations
in electron tunneling. However, the present study has revealed that
this physical simulation is not valid for a larger surface gap, which
is the energy gap of the insulator's surface states. To address this
issue, a refined pairing term that depends on the surface gap has
been obtained as a second-order effect of the proximity effect, whereas
the lowest order produces a constant pairing strength. By carefully
considering the dependence of pairing strength on the surface gap,
the Majorana phase diagram is re-achieved and a significant difference
from previous work is observed, where the pairing strength was assumed
to be independent of the surface gap and resulted in a conical phase
boundary.
\end{abstract}
\maketitle

\section{Introduction}

The zero-energy state of quasi-excitation in a hybrid system, known
as Majorana zero mode, has been used to physically simulate the Majorana
fermion \citep{majorana1937teoria,Kitaev_2001,Fu_ling_2008,Lutchyn_2010,Oreg_2010,Qi_2010,Alicea_2012,Mourik_2012,Wang_2015,Lutchyn2018}.
Over the last two decades, two common proposals for simulating Majorana
fermions have been developed based on hybrid condensed matter systems:
the topological insulators (TI) \citep{Hasan&Kane_2010,Fu&Kane_2007,moore2007topological,Hsieh2008topological,3D_TI_experimental_2009,Modeling_2010}
in proximity to an \textit{s}-wave superconductor (TI-SC) \citep{Fu_ling_2008,Qi_2010}
and the nanowire with spin-orbit coupling in contact with an \textit{s}-wave
superconductor \citep{Lutchyn_2010,Oreg_2010}. A Kitaev-type pairing
term on the surface of TI or nanowire can be induced by the proximity
effect \citep{DeGennes1964}, which is a virtual process of exchanging
the quasi-excitation in the superconductor.

In previous theoretical studies \citep{Fu_ling_2008,Qi_2010,Chung_2011,Wang_2015},
the Majorana zero mode was predicted when the Kitaev-type pairing
was always induced by the proximity effect in the lowest order. Therefore,
whether the pairing can be effectively induced to lead to a topological
phase transition depends on certain conditions. Moreover, there have
been controversies about the existence of chiral Majorana fermion
modes in some experiments, including a retracted paper \citep{He2017chiral}
that supported the lower-order theory, but was later disproved by
another experiment \citep{2020Absence}. To resolve these issues,
a more precise low-energy effective theory with higher-order proximity
effect was considered by the conventional Fröhlich-Nakajima (Schrieffer-Wolff)
transformation \citep{S.Nakajima_1955,Schrieffer&Wolff_1966,Frohlich_1950,Qiao_2022}
to refine the phase diagram .

Our study finds that the pairing strength depends on the surface gap
$m$, which is the energy gap of the surface states of the insulator,
in comparison with the pairing strength that was previously thought
to be independent of $m$. We take into account the surface gap dependence
and achieve a closed topological phase diagram in $m-\mu-\Delta$
space, where $\mu$ represents the chemical potential and $\Delta$
represents the constant pairing strength. This closed phase diagram
is different from the previous open phase diagram of the conic shape
\citep{Qi_2010}. Lastly, we note that the pairing strength becomes
divergent as the magnitude of surface gap $|m|$ approaches the superconducting
gap.

\section{Low-energy Effective Hamiltonian for TI-SC system}

Unlike ordinary insulators, topological insulators (TIs) exhibit surface
states in the vicinity of the Fermi level, which can be described
by the two-dimensional massless Dirac Hamiltonian $\mathcal{H}_{\mathrm{surf}}=v(k_{x}\sigma_{x}+k_{y}\sigma_{y})$,
with momentum space representation $\boldsymbol{\varphi}_{\mathbf{k}}=[\varphi_{\mathbf{k}\uparrow},\varphi_{\mathbf{k}\downarrow}]^{T}$
\citep{Modeling_2010,SurfaceObservation_2009,WenXiaoGang_2008}. Here,
$v$ denotes the Fermi velocity and $\sigma_{x},\sigma_{y}$ are the
Pauli matrices. By doping the TI material with magnetic elements such
as Fe or Cr, a mass term $m\text{\ensuremath{\sigma_{z}}}$ can be
introduced \citep{RuiYu_2010}, which opens a band gap of $2m$ for
the surface state. The magnitude of the mass $m$ (hereafter referred
to as the surface gap) depends on the magnetic ordering structure
and can be tuned by an external magnetic field. In reality, the low-energy
physics of TI thin films is more accurately described by a four-band
model, which includes tunneling between the opposite surfaces \citep{Wang_2015,RuiYu_2010}.
For the sake of clarity, we employ a reduced two-band Hamiltonian
that captures the essential features of TIs: $H_{\mathrm{TI}}=\int d^{2}k/(2\pi)^{2}\,\boldsymbol{\varphi}_{\mathbf{k}}^{\dagger}\cdot\mathcal{H}_{\mathbf{k}}\cdot\boldsymbol{\varphi}_{\mathbf{k}}$
\citep{Qi_2010,Chung_2011}, where the Hamiltonian matrix is given
by 
\begin{equation}
\mathcal{H}_{\mathbf{k}}=(m+m_{1}k^{2})\sigma_{z}-\mu+v(k_{x}\sigma_{x}+k_{y}\sigma_{y}).\label{eq:Ti-Hamiltonian}
\end{equation}
Here, $\mu$ is the chemical potential and $m_{1}k^{2}:=m_{1}(k_{x}^{2}+k_{y}^{2})$
is the parabolic band component, which is crucial for determining
the topological properties \citep{Qi_2010}. The TI thin film is placed
in contact with an \textit{s}-wave superconductor, which is described
by the BCS Hamiltonian

\begin{align}
\begin{aligned}H_{\mathrm{SC}}\end{aligned}
= & \int\frac{d^{3}k}{(2\pi)^{3}}\:\mathbf{c}_{\mathbf{K}}^{\dagger}[\epsilon_{s}\sigma_{z}+\Delta_{s}\sigma_{x}]\mathbf{c}_{\mathbf{K}},\label{eq:SC}
\end{align}
where $\mathbf{c}_{\mathbf{K}}=[c_{\mathbf{K}\uparrow},c_{-\mathbf{K}\downarrow}^{\dagger}]^{T}$
represents the Nambu spinor, $\Delta_{s}$ is the superconducting
gap, and $\epsilon_{s}=K^{2}/(2m_{s})-\mu_{s}$ is the the kinetic
energy of electron above the Fermi level $\mu_{s}$. The interaction
between the TI surface and the superconductor is modeled by the electron
tunneling Hamiltonian:
\begin{equation}
H_{\mathrm{T}}=J\int\frac{d^{3}k}{(2\pi)^{3}}\sum_{\sigma=\uparrow,\downarrow}[c_{\mathbf{K}\sigma}\varphi_{\mathbf{k}\sigma}^{\dagger}+\text{ H. c. }].\label{eq:tunneling-H}
\end{equation}
Here, we have assumed that the momentum paralleled to the surface
of TIs ($\mathbf{k}=\mathbf{K}_{//}\equiv(k_{x},k_{y}),\mathbf{K}_{\perp}\equiv k_{z}$)
and the spin are conserved during the electron tunneling process.
To eliminate virtual processes in the electron exchange between the
TI surface and the superconductor, we apply the Fröhlich-Nakajima
(Schrieffer-Wolff) transformation, resulting in an effective low-energy
Hamiltonian for the TI-SC system (see Appendix \ref{sec:low-energy-H})

\begin{equation}
H_{\mathrm{eff}}\simeq\int\frac{d^{2}k}{(2\pi)^{2}}\,\boldsymbol{\varphi}_{\mathbf{k}}^{\dagger}\cdot[(1-\frac{\tilde{\Delta}}{\Delta_{s}})\mathcal{H}_{\mathbf{k}}+\tilde{\Delta}\sigma_{x}]\cdot\boldsymbol{\varphi}_{\mathbf{k}}.\label{eq:H_eff}
\end{equation}
Here, the pairing strength $\tilde{\Delta}$ induced by the SC proximity
effect is

\begin{equation}
\tilde{\Delta}\simeq\left(1-(\frac{m}{\Delta_{s}})^{2}\right)^{-\frac{1}{2}}\Delta,\label{eq:Delta}
\end{equation}
where $\Delta\text{:}=J^{2}\sqrt{m_{s}/(2\mu_{s})}$ is the constant
pairing strength. Notably, the pairing strength $\tilde{\Delta}$
depends on the ratio of the surface gap and the superconducting gap,
i.e., $m/\Delta_{s}$. Moreover, there is an overall correction term
proportional to $\tilde{\Delta}/\Delta_{s}$ for original TI Hamiltonian.
As a result, the renormalized surface gap and the chemical potential
of TI become, respectively,

\begin{equation}
\frac{\tilde{m}}{m}=\frac{\tilde{\mu}}{\mu}=(1-\frac{\tilde{\Delta}}{\Delta_{s}}),\label{eq:Renorm}
\end{equation}

This parametric dependence of the pairing strength $\tilde{\Delta}$
on the surface gap $m$ leads to a significant change in the topological
phase diagram, as will be discussed in detail in the next section.

\section{Chern Number and Phase Diagram}

Typically, the parameter space is partitioned into distinct regions
based on the topological invariant, such as the Chern number, to obtain
the phase diagram with the topological phases located in those regions.
In the case of the hybrid system described by the low-energy effective
Hamiltonian $H_{\mathrm{eff}}$, the Chern number can be evaluated
as follows:

\begin{equation}
\mathcal{N}=\left\{ \begin{array}{cc}
0, & \tilde{m}>\sqrt{\tilde{\Delta}^{2}+\tilde{\mu}^{2}}\\
1, & -\sqrt{\tilde{\Delta}^{2}+\tilde{\mu}^{2}}<\tilde{m}<\sqrt{\tilde{\Delta}^{2}+\tilde{\mu}^{2}}\\
2, & \tilde{m}<-\sqrt{\tilde{\Delta}^{2}+\tilde{\mu}^{2}}
\end{array}\right..\label{eq:Chern}
\end{equation}
It follows from Eq. (\ref{eq:Chern}) that the topological phase transition
occurs at the condition $\tilde{m}^{2}=\tilde{\Delta}^{2}+\tilde{\mu}^{2}$,
which seems the same as the formerly results: $m^{2}=\Delta^{2}+\mu^{2}$
in ref. \citep{Qi_2010,Chung_2011}. However, the renormalized surface
gap $\tilde{m}$, the chemical potential $\tilde{\mu}$ and the induced
pairing $\tilde{\Delta}$ all depend on the surface gap $m$, which
is different from the formerly results where the three parameters
are independent of each other.

\begin{figure*}[t]
\noindent \begin{centering}
\includegraphics[width=15cm]{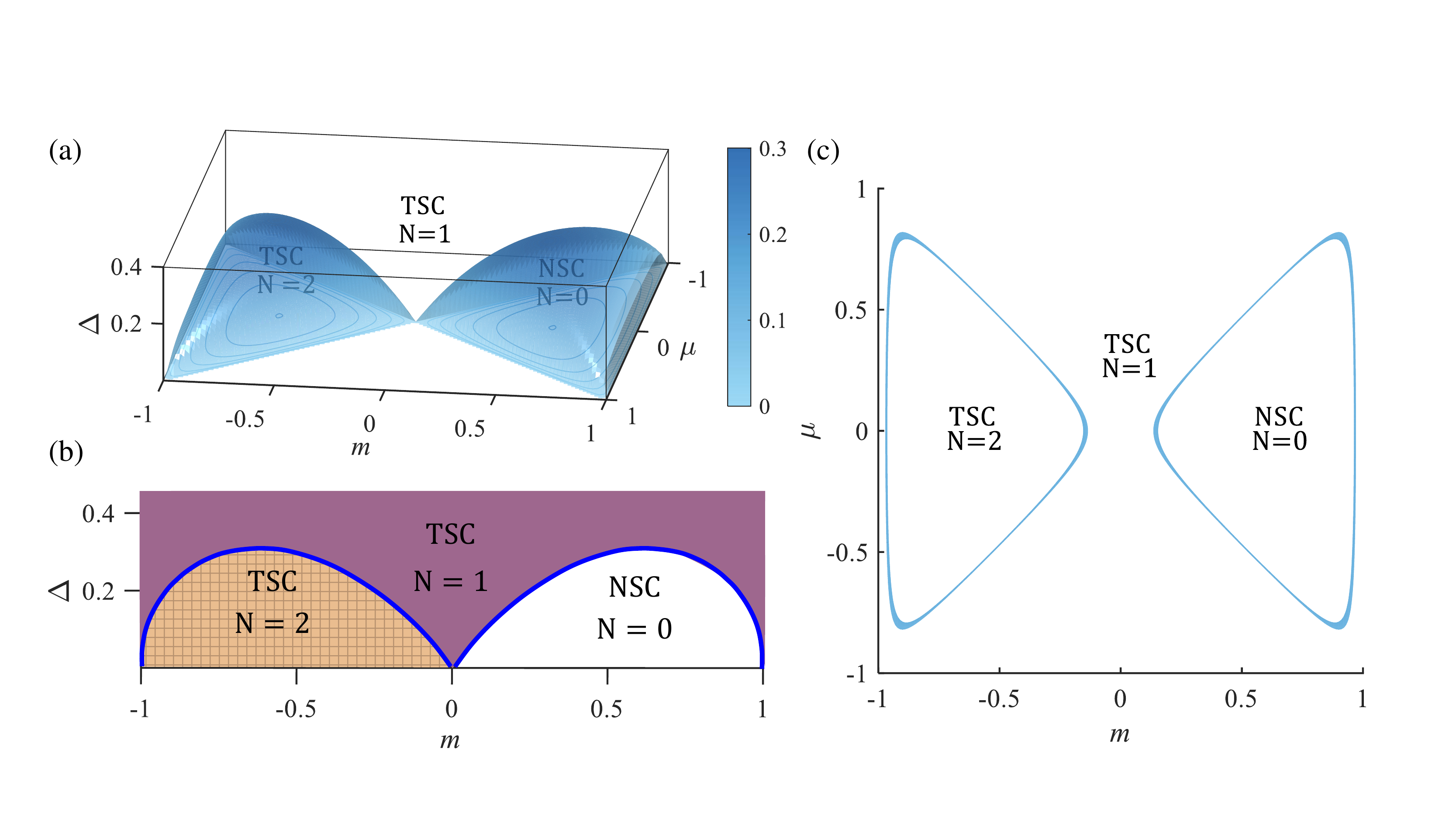}
\par\end{centering}
\noindent \raggedright{}FIG. 1. (a) Phase diagram of the TI-SC system
in $m-\mu-\Delta$ space (the coordinate axes are in units of $\Delta_{s}$).
The Chern number $\mathcal{N}=0$ corresponds to normal superconductor
(NSC) state, while $\mathcal{N}=1,2$ correspond to the topological
superconductor (TSC) state. The Chern number $\mathcal{N}$ equals
the number of Majorana edge modes. The region where $|m|>\Delta_{s}$
remains unclear. (b) Phase diagram of the TI-SC system for $\mu=0$.
(c) Phase diagram of the TI-SC system when $\Delta=0.125\Delta_{s}$.
\end{figure*}

By combining Eqs. (\ref{eq:Delta}, \ref{eq:Renorm}, \ref{eq:Chern}),
we can obtain the the phase diagram of the TI-SC system in $m-\mu-\Delta$
space, as presented in Fig. 1(a). The phase boundry exhibits a double-peak
structure that divides the phase space into three regions with distinct
Chern numbers. The left peak $(m<0)$ has a Chern number of $\mathcal{N}=2$,
while the right peak has $\mathcal{N}=0$. Outside of the peaks, the
Chern number is $\mathcal{N}=1$. Next, we compare the phase boundary
for $\mu=0$, as illustrated in Fig. 1(b), with previous works on
the $m-\Delta$ phase diagram that has a linear phase boundary described
by $\Delta=\pm m$ \citep{Qi_2010}. Additionally, we present the
phase boundary when $\Delta=0.125\Delta_{s}$ in Fig. 1(c). It is
worth noting that as the constant pairing $\Delta$ increases, the
region that corresponds to $\mathcal{N}=2$ and $\mathcal{N}=0$ becomes
smaller, eventually shrinking to a point when $\Delta=0.30\Delta_{s}$,
which corresponds to the maximum value of $\Delta$ on the phase boundry
curve in Fig. 1(b). Beyond this value (i.e., when $\Delta>0.30\Delta_{s}$),
the Chern number will always be 1, and no phase transition will occur
when $m$ is adjusted.

It is important to note that the pairing strength $\tilde{\Delta}$
becomes divergent when the surface gap $|m|$ approaches the superconducting
gap $\text{\ensuremath{\Delta_{s}}}$. Because of the divergent behavior,
the boundary sharply gets closed when $|m|$ is near $\Delta_{s}$.
The divergence indicates that the perturbation theory fails when $|m|$
is approximately equal to or greater than $\Delta_{s}$. In such cases,
it remains an open question whether an effective Hamiltonian can be
used to describe the hybrid system. When the surface gap is sufficiently
small ($|m|\ll\Delta_{s}$), the pairing strength becomes constant,
and the $m-\mu-\Delta$ phase diagram reverts to the corn form presented
in previous research \citep{Qi_2010}, resulting in a linear phase
boundary for $\mu=0$. Meanwhile, the $m-\mu$ phase diagram returns
to a parabolic curve, as described in previous studies of nanowire-superconductor
systems \citep{Alicea_2012} (the surface gap $m$ in TI system will
correspond to the external magnetic field $B$ in nanowire system)

The refined phase diagram shown in Fig. 1(b), in comparison with the
previous one \citep{Qi_2010}, reveals a smaller range of values for
the topological Majorana phase (Chern number $\mathcal{N}=1$). Therefore,
it is crucial to compare the surface gap of TI materials with the
superconducting gap in experiments to simulating Majorana fermions.
Currently, the superconducting gaps of commonly used materials for
Majorana detection, such as Nb, Al, and $\mathrm{NbSe_{2}}$, are
1.5, 2.0, and 2.15 meV, respectively \citep{He2017chiral,2020Absence,XueQiKun2015,clayman1971superconducting}.
While the surface gap of TI thin films is typically of a larger order
than the superconducting gap, such as a first-principles calculation
of the Hall conductance for Fe-doped $\mathrm{Bi_{2}Se_{3}}$, which
shows that the surface gap for 3, 4, and 5 quintuple layers are 90,
42, and 21 meV, respectively \citep{RuiYu_2010}. Moreover, in another
experimental work using angle-resolved photoemission spectroscopy,
the measured surface gap of $\mathrm{Bi_{2}Se_{3}}$ doped with 1\%
Mn is 7 meV \citep{2010massive}. Both of these data suggest that
the surface gap may exceed the superconducting gap in actual experiments,
raising questions about the effectiveness of simulating Majorana fermions
in these systems.

\section{Conclusion}

After re-examining the preconditions of the effective Hamiltonian
for a topological insulator (TI) in proximity to an \textsl{s}-wave
superconductor, we have found that the pairing strength, denoted as
$\tilde{\Delta}$, is dependent on the surface gap $m$ of TI to a
higher-order than previously assumed. By considering this higher-order
proximity effect, we have achieved a refined topological phase diagram
in the $m-\mu-\Delta$ space ($\mu$ represents the chemical potential
and $\Delta$ represents the constant pairing term), which is different
from the conic shape phase boundary resulting from the lower-order
approximation used in previous literature \citep{Qi_2010}. Moreover,
we have found that the pairing strength $\tilde{\Delta}$ becomes
divergent as $|m|$ approaches the superconducting gap $\Delta_{s}$.
Therefore, the validity of the effective Hamiltonian and conductance
signatures for Majorana fermion detection are only credible when $|m|\text{<}\Delta_{s}$,
and the confidence of the results deteriorates as $|m|$ is closer
to $\Delta_{s}$. Therefore, the Majorana fermion simulation for TI-SC
system should be realized with a modest surface gap, $|m|\text{<}\Delta_{s}$.
However, the existing experimental data suggests that the surface
gap may be beyond the effective range given by our refined phase,
and we are eagerly awaiting a response to this problem.

\appendix
\begin{widetext}

\section{The low-energy effective Hamiltonian of topological insulator-superconductor
system\label{sec:low-energy-H}}

In the appendix, we derive the low-energy effective Hamiltonian (\ref{eq:H_eff})
for the topological insulator (TI) in proximity to an \textsl{s}-wave
superconductor (SC) system by the Fröhlich-Nakajima (Schrieffer-Wolff)
transformation. The total Hamiltonian for the TI-SC system includes
three main terms: $H=H_{\mathrm{TI}}+H_{\mathrm{SC}}+H_{\mathrm{T}}\equiv H_{0}+H_{1}$.
And the surface state of a magnetic TI thin film can be described
by a two band model:
\begin{equation}
\begin{aligned}H_{\mathrm{TI}} & =\int\frac{d^{2}k}{(2\pi)^{2}}\Bigl\{\left(m_{k}-\mu\right)\varphi_{\mathbf{k}\uparrow}^{\dagger}\varphi_{\mathbf{k}\uparrow}-\left(m_{k}+\mu\right)\varphi_{\mathbf{k}\downarrow}^{\dagger}\varphi_{\mathbf{k}\downarrow}+\left[v(k_{x}+ik_{y})\varphi_{\mathbf{k}\downarrow}^{\dagger}\varphi_{\mathbf{k}\uparrow}+\mathrm{H.c.}\right]\Bigr\}\end{aligned}
\end{equation}
with $\varphi_{\mathbf{k}\sigma}$ annihilating an electron of momentum
$\mathbf{k}$ and spin $\sigma=\uparrow,\downarrow$. Here, $m_{k}=m+m_{1}(k_{x}^{2}+k_{y}^{2})$
is the mass term with the surface gap $m$, $\mu$ is the chemical
potential and $v$ is the Fermi velocity. The \textsl{s}-wave SC providing
the superconducting proximity effect for the TI film is described
by the BCS Hamiltonian (under self-consistent field approximation)

\begin{align}
\begin{aligned}H_{\mathrm{SC}}=\int\frac{d^{3}k}{(2\pi)^{3}}\biggl[\epsilon_{s}\left(c_{\mathbf{K}\uparrow}^{\dagger}c_{\mathbf{K}\uparrow}+c_{-\mathbf{K}\downarrow}^{\dagger}c_{-\mathbf{K}\downarrow}\right)\end{aligned}
 & +\Delta_{s}\left(c_{\mathbf{K}\uparrow}^{\dagger}c_{-\mathbf{K}\downarrow}^{\dagger}+\text{ H. c }\right)\biggr],
\end{align}
with the superconducting gap $\Delta_{s}$ and the kinetic energy
of electron $\epsilon_{s}=K^{2}/(2m_{s})-\mu_{s}$ above the Fermi
level $\mu_{s}$. The tunneling interaction between TI and SC by the
contact plane $z=0$ reads as:

\begin{align}
H_{1} & =-J\sum_{\sigma=\uparrow,\downarrow}\iint dxdy\left[\varphi_{\sigma}^{\dagger}(x,y)c_{\sigma}(x,y,0)+\text{ H. c. }\right].\label{eq:tunneling-H-1}
\end{align}
Here, $J$ is the tunneling strength, and $\varphi_{\sigma}(\mathbf{x})$,
$c_{\sigma}(\mathbf{X})$ are the inverse Fourier transforms of $\varphi_{\mathbf{k}\sigma}$,
$c_{\mathbf{K}\sigma}$: 
\begin{equation}
\varphi(\mathbf{x})=\int\frac{d^{2}k}{(2\pi)^{2}}\varphi_{\mathbf{k}\sigma}e^{\mathrm{i}\mathbf{k\cdot x}},\quad c(\mathbf{X})=\int\frac{d^{3}k}{(2\pi)^{3}}c_{\mathbf{K}\sigma}e^{\mathrm{i}\mathbf{K\cdot X}},\label{eq:inverse-FT}
\end{equation}
Notice that $\mathbf{x,k}$ are the two-dimensional component (parallel
to TI surface) of $\mathbf{X,K}$ respectively. By the Fourier transformation,
we can rewrite the tunneling Hamiltonian (\ref{eq:tunneling-H-1})
in momentum space:

\begin{equation}
\begin{aligned}H_{1}=-J\int\frac{d^{3}k}{(2\pi)^{3}}\left(\varphi_{\mathbf{k}\uparrow}^{\dagger}c_{\mathbf{K}\uparrow}+\varphi_{\mathbf{k}\downarrow}^{\dagger}c_{\mathbf{K}\downarrow}+\text{ H. c.}\right)\end{aligned}
,\label{eq:tunneling-interaction}
\end{equation}

To describe the quasi-particles in SC, we introduce the Bogoliubov
transformation as follows

\begin{equation}
\begin{aligned}\eta_{\mathbf{K}\uparrow} & :=\cos\theta_{\mathbf{K}}c_{\mathbf{K}\uparrow}+\sin\theta_{\mathbf{K}}c_{-\mathbf{K}\downarrow}^{\dagger},\\
\eta_{-\mathbf{K}\downarrow}^{\dagger} & :=-\sin\theta_{\mathbf{K}}c_{\mathbf{K}\uparrow}+\cos\theta_{\mathbf{\mathbf{K}}}c_{-\mathbf{K}\downarrow}^{\dagger},
\end{aligned}
\end{equation}
with $\tan2\theta_{\mathbf{K}}=\Delta_{s}/\epsilon_{s}$. Then the
BCS Hamiltonian can be diagonalized as

\begin{equation}
\begin{gathered}\begin{aligned}H_{\mathrm{SC}}=\int\frac{d^{3}k}{(2\pi)^{3}}E_{s}\left(\eta_{\mathbf{K}\uparrow}^{\dagger}\eta_{\mathbf{K}\uparrow}+\eta_{-\mathbf{K}\downarrow}^{\dagger}\eta_{-\mathbf{K}\downarrow}\right)\end{aligned}
,\end{gathered}
\end{equation}
where $E_{s}=\sqrt{\epsilon_{s}^{2}+\Delta_{s}^{2}}$ is the energy
spectrum of the quasi-partical in SC.

Similarly, we rewrite the tunneling interaction (\ref{eq:tunneling-interaction})
with Bogoliubov quasi-particle operators $\eta_{\mathbf{K}\sigma}$
as

\begin{align}
H_{1} & =-J\int\frac{d^{3}k}{(2\pi)^{3}}\eta_{\mathbf{K}\uparrow}\left(-\cos\theta_{\mathbf{K}}\varphi_{\mathbf{k}\uparrow}^{\dagger}+\sin\theta_{\mathbf{K}}\varphi_{-\mathbf{k}\downarrow}\right)+\eta_{\mathbf{K}\downarrow}\left(-\cos\theta_{\mathbf{K}}\varphi_{\mathbf{k}\downarrow}^{\dagger}-\sin\theta_{\mathbf{K}}\varphi_{-\mathbf{k}\uparrow}\right)\nonumber \\
 & +\eta_{\mathbf{K}\uparrow}^{\dagger}\left(-\sin\theta_{\mathbf{K}}\varphi_{-\mathbf{k}\downarrow}^{\dagger}+\cos\theta_{\mathbf{K}}\varphi_{\mathbf{k}\uparrow}\right)+\eta_{\mathbf{K}\downarrow}^{\dagger}\left(\sin\theta_{\mathbf{K}}\varphi_{-\mathbf{k}\uparrow}^{\dagger}+\cos\theta_{\mathbf{K}}\varphi_{\mathbf{k}\downarrow}\right).
\end{align}

Now, we apply the Fröhlich-Nakajima (Schrieffer-Wolff) transformation
to eliminate the quasi-excitation of the SC. Treating $H_{1}$ as
a perturbation term, we perform a canonical transformation $\mathrm{e}^{S}$
to the total Hamiltonian:

\begin{align}
H_{\mathrm{eff}}=\mathrm{e}^{S}H\mathrm{e}^{-S} & =H+[H,S]+\text{\ensuremath{\frac{1}{2!}}}[[H,S],S]+\ldots\nonumber \\
 & =H_{0}+(H_{1}+[H_{0},S])+\frac{1}{2}[(H_{1}+[H_{0},S]),S]+\frac{1}{2}[H_{1},S]+\ldots
\end{align}
Moreover, we require that the transformed Hamiltonian has no first
order term, i.e. $[H_{0},S]+H_{1}=0$, and the ansatz for the anti-Hermitian
transformation $S$ is set as:
\begin{equation}
\begin{aligned}S=\int\frac{d^{3}k}{(2\pi)^{3}} & \bigg\{\eta_{\uparrow\mathbf{K}}\left[A_{\mathbf{K}}\varphi_{\mathbf{k}\uparrow}^{\dagger}+B_{\mathbf{K}}\varphi_{-\mathbf{k}\downarrow}+E_{\mathbf{K}}\varphi_{\mathbf{k}\downarrow}^{\dagger}+F_{\mathbf{K}}\varphi_{-\mathbf{k}\uparrow}\right]\\
 & +\eta_{\uparrow\mathbf{K}}^{\dagger}\left[A_{\mathbf{K}}^{*}\varphi_{\mathbf{k}\uparrow}+B_{\mathbf{K}}^{{*}}\varphi_{-\mathbf{k}\downarrow}^{\dagger}+E_{\mathbf{K}}^{{*}}\varphi_{\mathbf{k}\downarrow}+F_{\mathbf{K}}^{{*}}\varphi_{-\mathbf{k}\uparrow}^{\dagger}\right]\\
 & +\eta_{\downarrow\mathbf{K}}\left[C_{\mathbf{K}}\varphi_{\mathbf{k}\downarrow}^{\dagger}+D_{\mathbf{K}}\varphi_{-\mathbf{k}\uparrow}+H_{\mathbf{K}}\varphi_{\mathbf{k}\uparrow}^{\dagger}+L_{\mathbf{K}}\varphi_{-\mathbf{k}\downarrow}\right]\\
 & +\eta_{\downarrow\mathbf{K}}^{\dagger}\left[C_{\mathbf{K}}^{{*}}\varphi_{\mathbf{k}\downarrow}+D_{\mathbf{K}}^{{*}}\varphi_{-\mathbf{k}\uparrow}^{\dagger}+H_{\mathbf{K}}^{{*}}\varphi_{\mathbf{k}\uparrow}+L_{\mathbf{K}}^{{*}}\varphi_{-\mathbf{k}\downarrow}^{\dagger}\right]\bigg\}.
\end{aligned}
\label{eq:transformation-S}
\end{equation}
By satisfying the condition $[H_{0},S]+H_{1}=0$, the undetermined
coefficients of the transformation $S$ in Eq. (\ref{eq:transformation-S})
can be obtained as

\begin{equation}
\begin{aligned} & A_{\mathbf{K}}=J\cos\theta_{\mathbf{K}}\frac{E_{s}+m_{k}-\mu}{\Pi_{-}} &  & E_{\mathbf{K}}=J\cos\theta_{\mathbf{K}}\frac{vk_{+}}{\Pi_{-}}\\
 & B_{\mathbf{K}}=-J\sin\theta_{\mathbf{K}}\frac{E_{s}+m_{k}-\mu}{\Pi_{+}} &  & F_{\mathbf{K}}=-J\sin\theta_{\mathbf{K}}\frac{vk_{+}}{\Pi_{+}}\\
 & C_{\mathbf{K}}=J\cos\theta_{\mathbf{K}}\frac{E_{s}-m_{k}+\mu}{\Pi_{-}} &  & H_{\mathbf{K}}=J\cos\theta_{\mathbf{K}}\frac{vk_{-}}{\Pi_{-}}\\
 & D_{\mathbf{K}}=J\sin\theta_{\mathbf{K}}\frac{E_{s}-m_{k}-\mu}{\Pi_{+}} &  & L_{\mathbf{K}}=J\sin\theta_{\mathbf{K}}\frac{vk_{-}}{\Pi_{+}}
\end{aligned}
\label{eq:Coef}
\end{equation}
with $\Pi_{\pm}\equiv\left(E_{s}\mp\mu\right)^{2}-m_{k}^{2}-v^{2}k^{2}$
and $k_{\pm}\equiv k_{x}\pm ik_{y}$. When the surface gap $m$ is
not too large (compared to the superconducting gap $\Delta_{s}$)
and the electron tunneling strength $J$ is weak (i.e. $|\Pi_{\pm}|\gg J$),
the effective Hamiltonian of the TI-SC in second-order perturbation
is further obtained as $H_{\mathrm{eff}}=H_{0}+\frac{1}{2}[H_{1},S]$,
where the second-order term can be calculated as

\begin{equation}
\begin{aligned}\frac{1}{2}[H_{1},S]= & \frac{J}{2}\int\frac{d^{3}k}{(2\pi)^{3}}\left(-\cos\theta_{\mathbf{K}}\varphi_{\mathbf{k}\uparrow}^{\dagger}+\sin\theta_{\mathbf{K}}\varphi_{\mathbf{-k}\downarrow}\right)\left(A_{\mathbf{K}}^{*}\varphi_{\mathbf{k}\uparrow}+B_{\mathbf{K}}^{{*}}\varphi_{-\mathbf{k}\downarrow}^{\dagger}+E_{\mathbf{K}}^{{*}}\varphi_{\mathbf{k}\downarrow}+F_{\mathbf{K}}^{{*}}\varphi_{-\mathbf{k}\uparrow}^{\dagger}\right)\\
 & +\left(-\sin\theta_{\mathbf{K}}\varphi_{\mathbf{-k}\downarrow}^{\dagger}+\cos\theta_{\mathbf{K}}\varphi_{\mathbf{k}\uparrow}\right)\left(A_{\mathbf{K}}\varphi_{\mathbf{k}\uparrow}^{\dagger}+B_{\mathbf{K}}\varphi_{\mathbf{-k}\downarrow}+E_{\mathbf{K}}\varphi_{\mathbf{k}\downarrow}^{\dagger}+F_{\mathbf{K}}\varphi_{\mathbf{k}\uparrow}\right)\\
 & +\left(-\cos\theta_{\mathbf{K}}\varphi_{\mathbf{k}\downarrow}^{\dagger}-\sin\theta_{\mathbf{K}}\varphi_{-\mathbf{k}\uparrow}\right)\left(C_{\mathbf{K}}^{{*}}\varphi_{\mathbf{k}\downarrow}+D_{\mathbf{K}}^{{*}}\varphi_{-\mathbf{k}\uparrow}^{\dagger}+H_{\mathbf{K}}^{{*}}\varphi_{\mathbf{k}\uparrow}+L_{\mathbf{K}}^{{*}}\varphi_{-\mathbf{k}\downarrow}^{\dagger}\right)\\
 & +\left(\sin\theta_{\mathbf{K}}\varphi_{-\mathbf{k}\uparrow}^{\dagger}+\cos\theta_{\mathbf{K}}\varphi_{\mathbf{k}\downarrow}\right)\left(C_{\mathbf{K}}\varphi_{\mathbf{k}\downarrow}^{\dagger}+D_{\mathbf{K}}\varphi_{-\mathbf{k}\uparrow}+H_{\mathbf{K}}\varphi_{\mathbf{k}\uparrow}^{\dagger}+L_{\mathbf{K}}\varphi_{-\mathbf{k}\downarrow}\right)\\
= & \frac{J}{2}\int\frac{d^{3}k}{(2\pi)^{3}}\Bigl\{\left[-2A_{\mathbf{K}}\cos\theta_{\mathbf{K}}+2D_{\mathbf{K}}\sin\theta_{\mathbf{K}}\right]\varphi_{\mathbf{k}\uparrow}^{\dagger}\varphi_{\mathbf{k}\uparrow}+\left[-2B_{\mathbf{K}}\sin\theta_{\mathbf{K}}-2C_{\mathbf{K}}\cos\theta_{\mathbf{K}}\right]\varphi_{\mathbf{k}\downarrow}^{\dagger}\varphi_{\mathbf{k}\downarrow}\\
 & +\left[(-2E_{\mathbf{K}}\cos\theta_{\mathbf{K}}+2F_{\mathbf{K}}\sin\theta_{\mathbf{K}})\varphi_{\mathbf{k}\downarrow}^{\dagger}\varphi_{\mathbf{k}\uparrow}+\mathrm{H.c.}\right]+\left[(E_{\mathbf{K}}\sin\theta_{\mathbf{K}}+F_{\mathbf{K}}\cos\theta_{\mathbf{K}})\text{(}\varphi_{\mathbf{k}\downarrow}^{\dagger}\varphi_{\mathbf{-k}\uparrow}^{\dagger}+\varphi_{\mathbf{k}\uparrow}\varphi_{-\mathbf{k}\uparrow})+\mathrm{H.c.}\right]\\
 & +\left[\left(A_{\mathbf{K}}\sin\theta_{\mathbf{K}}-B_{\mathbf{K}}\cos\theta_{\mathbf{K}}+C_{\mathbf{K}}\sin\theta_{\mathbf{K}}+D_{\mathbf{K}}\cos\theta_{\mathbf{K}}\right)\varphi_{\mathbf{k}\uparrow}^{\dagger}\varphi_{-\mathbf{k}\downarrow}+\mathrm{H.c.}\right]\Bigr\}
\end{aligned}
\label{eq:here}
\end{equation}

In the above calculation (\ref{eq:here}), we have utilized the relations
of the coefficients: $A_{\mathbf{K}}=A_{-\mathbf{K}}=A_{\mathbf{K}}^{*}$
and $E_{\mathbf{K}}=-E_{-\mathbf{K}}=H_{\mathbf{K}}^{{*}}$ and discarded
the superconducting terms $\eta^{\dagger}\eta^{\dagger}\text{, \ensuremath{\eta^{\dagger}\eta\text{, }\eta\eta}}$
due to the decoupled hybird system in the second-order approximation.
By substituting the coefficients in Eq. (\ref{eq:Coef}) into Eq.
(\ref{eq:here}), the effective Hamiltonian of TI dressed by the superconducting
proximity effect\textbf{ }by is obtained as

\begin{equation}
\begin{aligned}H_{\mathrm{eff}}=\int\frac{d^{2}k}{(2\pi)^{2}}\Bigl\{\tilde{m}_{k} & (\varphi_{\mathbf{k}\uparrow}^{\dagger}\varphi_{\mathbf{k}\uparrow}-\varphi_{\mathbf{k}\downarrow}^{\dagger}\varphi_{\mathbf{k}\downarrow})-\tilde{\mu}(\varphi_{\mathbf{k}\uparrow}^{\dagger}\varphi_{\mathbf{k}\uparrow}+\varphi_{\mathbf{k}\downarrow}^{\dagger}\varphi_{\mathbf{k}\downarrow})+[\tilde{v}(k_{x}+ik_{y})\varphi_{\mathbf{k}\downarrow}^{\dagger}\varphi_{\mathbf{k}\uparrow}+\mathrm{H.c.}]\\
 & \tilde{\Delta}_{\mathbf{k}}(\varphi_{\mathbf{k}\uparrow}^{\dagger}\varphi_{\mathbf{-k}\downarrow}^{\dagger}+\mathrm{H.c.})+[\tilde{\Lambda}_{\mathbf{k}}\text{(}\varphi_{\mathbf{k}\downarrow}^{\dagger}\varphi_{\mathbf{-k}\downarrow}^{\dagger}+\varphi_{\mathbf{k}\uparrow}\varphi_{-\mathbf{k}\uparrow})+\mathrm{H.c.}]\Bigr\},
\end{aligned}
\end{equation}
where the renormalized mass term $\tilde{m}_{k}$, the chemical potential
$\tilde{\mu}$, the Fermi velocity of TI $\tilde{v}$ and the induced
pairing terms with the same spin $\tilde{\Delta}_{\mathbf{k}}$ and
the opposite spin $\tilde{\Lambda}_{\mathbf{k}}$ are respectively

\begin{equation}
\begin{aligned} & \tilde{m}_{k}=\left[1-J^{2}\int\frac{dk_{z}}{2\pi}(\frac{\cos^{2}\theta_{\mathbf{K}}}{\Pi_{-}}+\frac{\sin^{2}\theta_{\mathbf{K}}}{\Pi_{+}})\right]m_{k}\approx(1-J^{2}\int\frac{dk_{z}}{2\pi}\frac{1}{\Pi})m_{k},\\
 & \tilde{\mu}=\mu+J^{2}\int\frac{dk_{z}}{2\pi}\left[\frac{\cos^{2}\theta_{\mathbf{K}}\left(E_{s}+\mu\right)}{\Pi_{-}}-\frac{\sin^{2}\theta_{\mathbf{K}}\left(E_{s}-\mu\right)}{\Pi_{+}}\right]\approx\mu+J^{2}\int\frac{dk_{z}}{2\pi}\frac{\epsilon_{s}-\mu}{\Pi},\\
 & \tilde{v}=\left[1-J^{2}\int\frac{dk_{z}}{2\pi}\left(\frac{\cos^{2}\theta_{\mathbf{K}}}{\Pi_{-}}+\frac{\sin^{2}\theta_{\mathbf{K}}}{\Pi_{+}}\right)\right]v\text{\ensuremath{\approx(1-J^{2}\int\frac{dk_{z}}{2\pi}\frac{1}{\Pi})v},}\\
 & \tilde{\Delta}_{\mathbf{k}}=\frac{J^{2}}{2}\int\frac{dk_{z}}{2\pi}\sin2\theta_{\mathbf{K}}\left(\frac{E_{s}+\mu}{\Pi_{-}}+\frac{E_{s}-\mu}{\Pi_{+}}\right)\approx J^{2}\Delta_{s}\int\frac{dk_{z}}{2\pi}\frac{1}{\Pi},\\
 & \tilde{\Lambda}_{\mathbf{k}}=\frac{J^{2}}{4}\int\frac{dk_{z}}{2\pi}\sin2\theta_{\mathbf{K}}\left(\frac{1}{\Pi_{-}}-\frac{1}{\Pi_{+}}\right)vk_{+}\approx0.
\end{aligned}
\label{eq:renormalized-coff}
\end{equation}
with $\Pi\equiv E_{s}^{2}-m_{k}^{2}-v^{2}k^{2}$. In the above simplification
of (\ref{eq:renormalized-coff}), we have considered that the chemical
potential is much smaller than the SC gap (i.e. $\mu\ll\Delta_{s}$),
so the higher orders of $\mu/E_{s}$ are ignored. Besides, the renormalized
chemical potential $\tilde{\mu}$ is obtained by considering the small
mass term and the low-energy state of TI ($m_{k},vk\ll E_{s}$) simultaneously.

Finally, we need to finish the calculation of the two integrals that
remain in Eq. (\ref{eq:renormalized-coff}), one is:

\begin{equation}
\begin{aligned}\int\frac{dk_{z}}{2\pi}\frac{1}{\Pi} & =\int\frac{dk_{z}}{2\pi}\frac{1}{E_{s}^{2}-m_{k}^{2}-v^{2}k^{2}}\approx\int\frac{dk_{z}}{2\pi}\frac{1}{\Delta_{s}^{2}+\left(\frac{k_{z}^{2}}{2m_{s}}-\mu_{s}\right)^{2}-m^{2}}\\
 & \approx\frac{1}{\pi}\int_{-\hbar\omega_{D}}^{\hbar\omega_{D}}d\epsilon\sqrt{\frac{m_{s}}{2\left(\epsilon+\mu_{s}\right)}}\frac{1}{\Delta_{s}^{2}-m^{2}+\epsilon^{2}}\\
 & \approx\sqrt{\frac{m_{s}}{2\mu_{s}}}\frac{1}{\sqrt{\Delta_{s}^{2}-m^{2}}}.
\end{aligned}
\end{equation}

Above, we have considered that the TI is in the low energy state ($k^{2}$
is small) and employed the Debye truncation approximation (the Debye
frequency $\omega_{D}$). What's more, $\Delta_{s}\ll\omega_{D}\ll\mu_{s}$
has been used to simplify the result. Similarly, we can obtain the
other integral $\int\frac{dk_{z}}{2\pi}\frac{\epsilon_{s}}{\Pi}$$\approx$0,
then the effective Hamiltonian of (\ref{eq:H_eff}) in the main text
is finally achieved as

\begin{equation}
\begin{aligned}H_{\mathrm{eff}}=\int\frac{d^{2}k}{(2\pi)^{2}}\Bigl\{\tilde{m}_{k} & (\varphi_{\mathbf{k}\uparrow}^{\dagger}\varphi_{\mathbf{k}\uparrow}-\varphi_{\mathbf{k}\downarrow}^{\dagger}\varphi_{\mathbf{k}\downarrow})-\tilde{\mu}(\varphi_{\mathbf{k}\uparrow}^{\dagger}\varphi_{\mathbf{k}\uparrow}+\varphi_{\mathbf{k}\downarrow}^{\dagger}\varphi_{\mathbf{k}\downarrow})+[\tilde{v}(k_{x}+ik_{y})\varphi_{\mathbf{k}\downarrow}^{\dagger}\varphi_{\mathbf{k}\uparrow}+\mathrm{H.c.}]\\
 & \tilde{\Delta}\text{(\ensuremath{\varphi_{\mathbf{k}\uparrow}^{\dagger}\varphi_{\mathbf{-k}\downarrow}^{\dagger}}+\ensuremath{\mathrm{H.c.}})}\Bigr\},
\end{aligned}
\end{equation}
where the mass term, the chemical potential and the Fermi velocity
are renomalized by a universal factor

\begin{equation}
\begin{aligned}\frac{\tilde{m}_{k}}{m_{k}}=\frac{\tilde{\mu}}{\mu}=\frac{\tilde{v}}{v}\approx1-\tilde{\Delta}/\Delta_{s},\,\tilde{\Delta}\approx J^{2}\sqrt{\frac{m_{s}}{2\mu_{s}}}\left(1-(\frac{m}{\Delta_{s}})^{2}\right)^{-\frac{1}{2}}.\end{aligned}
\end{equation}

\end{widetext}

\bibliographystyle{apsrev4-2}
\bibliography{bibtex}

\begin{thebibliography}{30}%
\makeatletter
\providecommand \@ifxundefined [1]{%
 \@ifx{#1\undefined}
}%
\providecommand \@ifnum [1]{%
 \ifnum #1\expandafter \@firstoftwo
 \else \expandafter \@secondoftwo
 \fi
}%
\providecommand \@ifx [1]{%
 \ifx #1\expandafter \@firstoftwo
 \else \expandafter \@secondoftwo
 \fi
}%
\providecommand \natexlab [1]{#1}%
\providecommand \enquote  [1]{``#1''}%
\providecommand \bibnamefont  [1]{#1}%
\providecommand \bibfnamefont [1]{#1}%
\providecommand \citenamefont [1]{#1}%
\providecommand \href@noop [0]{\@secondoftwo}%
\providecommand \href [0]{\begingroup \@sanitize@url \@href}%
\providecommand \@href[1]{\@@startlink{#1}\@@href}%
\providecommand \@@href[1]{\endgroup#1\@@endlink}%
\providecommand \@sanitize@url [0]{\catcode `\\12\catcode `\$12\catcode
  `\&12\catcode `\#12\catcode `\^12\catcode `\_12\catcode `\%12\relax}%
\providecommand \@@startlink[1]{}%
\providecommand \@@endlink[0]{}%
\providecommand \url  [0]{\begingroup\@sanitize@url \@url }%
\providecommand \@url [1]{\endgroup\@href {#1}{\urlprefix }}%
\providecommand \urlprefix  [0]{URL }%
\providecommand \Eprint [0]{\href }%
\providecommand \doibase [0]{https://doi.org/}%
\providecommand \selectlanguage [0]{\@gobble}%
\providecommand \bibinfo  [0]{\@secondoftwo}%
\providecommand \bibfield  [0]{\@secondoftwo}%
\providecommand \translation [1]{[#1]}%
\providecommand \BibitemOpen [0]{}%
\providecommand \bibitemStop [0]{}%
\providecommand \bibitemNoStop [0]{.\EOS\space}%
\providecommand \EOS [0]{\spacefactor3000\relax}%
\providecommand \BibitemShut  [1]{\csname bibitem#1\endcsname}%
\let\auto@bib@innerbib\@empty
\bibitem [{\citenamefont {Majorana}(1937)}]{majorana1937teoria}%
  \BibitemOpen
  \bibfield  {author} {\bibinfo {author} {\bibfnamefont {E.}~\bibnamefont
  {Majorana}},\ }\href@noop {} {\bibfield  {journal} {\bibinfo  {journal}
  {Nuovo Cim.}\ }\textbf {\bibinfo {volume} {14}},\ \bibinfo {pages} {171}
  (\bibinfo {year} {1937})}\BibitemShut {NoStop}%
\bibitem [{\citenamefont {Kitaev}(2001)}]{Kitaev_2001}%
  \BibitemOpen
  \bibfield  {author} {\bibinfo {author} {\bibfnamefont {A.~Y.}\ \bibnamefont
  {Kitaev}},\ }\href {https://doi.org/10.1070/1063-7869/44/10s/s29} {\bibfield
  {journal} {\bibinfo  {journal} {Phy. Usp.}\ }\textbf {\bibinfo {volume}
  {44}},\ \bibinfo {pages} {131} (\bibinfo {year} {2001})}\BibitemShut
  {NoStop}%
\bibitem [{\citenamefont {Fu}\ and\ \citenamefont {Kane}(2008)}]{Fu_ling_2008}%
  \BibitemOpen
  \bibfield  {author} {\bibinfo {author} {\bibfnamefont {L.}~\bibnamefont
  {Fu}}\ and\ \bibinfo {author} {\bibfnamefont {C.~L.}\ \bibnamefont {Kane}},\
  }\href {https://doi.org/10.1103/PhysRevLett.100.096407} {\bibfield  {journal}
  {\bibinfo  {journal} {Phys. Rev. Lett.}\ }\textbf {\bibinfo {volume} {100}},\
  \bibinfo {pages} {096407} (\bibinfo {year} {2008})}\BibitemShut {NoStop}%
\bibitem [{\citenamefont {Lutchyn}\ \emph {et~al.}(2010)\citenamefont
  {Lutchyn}, \citenamefont {Sau},\ and\ \citenamefont
  {Das~Sarma}}]{Lutchyn_2010}%
  \BibitemOpen
  \bibfield  {author} {\bibinfo {author} {\bibfnamefont {R.~M.}\ \bibnamefont
  {Lutchyn}}, \bibinfo {author} {\bibfnamefont {J.~D.}\ \bibnamefont {Sau}},\
  and\ \bibinfo {author} {\bibfnamefont {S.}~\bibnamefont {Das~Sarma}},\ }\href
  {https://doi.org/10.1103/PhysRevLett.105.077001} {\bibfield  {journal}
  {\bibinfo  {journal} {Phys. Rev. Lett.}\ }\textbf {\bibinfo {volume} {105}},\
  \bibinfo {pages} {077001} (\bibinfo {year} {2010})}\BibitemShut {NoStop}%
\bibitem [{\citenamefont {Oreg}\ \emph {et~al.}(2010)\citenamefont {Oreg},
  \citenamefont {Refael},\ and\ \citenamefont {von Oppen}}]{Oreg_2010}%
  \BibitemOpen
  \bibfield  {author} {\bibinfo {author} {\bibfnamefont {Y.}~\bibnamefont
  {Oreg}}, \bibinfo {author} {\bibfnamefont {G.}~\bibnamefont {Refael}},\ and\
  \bibinfo {author} {\bibfnamefont {F.}~\bibnamefont {von Oppen}},\ }\href
  {https://doi.org/10.1103/PhysRevLett.105.177002} {\bibfield  {journal}
  {\bibinfo  {journal} {Phys. Rev. Lett.}\ }\textbf {\bibinfo {volume} {105}},\
  \bibinfo {pages} {177002} (\bibinfo {year} {2010})}\BibitemShut {NoStop}%
\bibitem [{\citenamefont {Qi}\ \emph {et~al.}(2010)\citenamefont {Qi},
  \citenamefont {Hughes},\ and\ \citenamefont {Zhang}}]{Qi_2010}%
  \BibitemOpen
  \bibfield  {author} {\bibinfo {author} {\bibfnamefont {X.-L.}\ \bibnamefont
  {Qi}}, \bibinfo {author} {\bibfnamefont {T.~L.}\ \bibnamefont {Hughes}},\
  and\ \bibinfo {author} {\bibfnamefont {S.-C.}\ \bibnamefont {Zhang}},\ }\href
  {https://doi.org/10.1103/PhysRevB.82.184516} {\bibfield  {journal} {\bibinfo
  {journal} {Phys. Rev. B}\ }\textbf {\bibinfo {volume} {82}},\ \bibinfo
  {pages} {184516} (\bibinfo {year} {2010})}\BibitemShut {NoStop}%
\bibitem [{\citenamefont {Alicea}(2012)}]{Alicea_2012}%
  \BibitemOpen
  \bibfield  {author} {\bibinfo {author} {\bibfnamefont {J.}~\bibnamefont
  {Alicea}},\ }\href {https://doi.org/10.1088/0034-4885/75/7/076501} {\bibfield
   {journal} {\bibinfo  {journal} {Rep. Prog. Phys.}\ }\textbf {\bibinfo
  {volume} {75}},\ \bibinfo {pages} {076501} (\bibinfo {year}
  {2012})}\BibitemShut {NoStop}%
\bibitem [{\citenamefont {Mourik}\ \emph {et~al.}(2012)\citenamefont {Mourik},
  \citenamefont {Zuo}, \citenamefont {Frolov}, \citenamefont {Plissard},
  \citenamefont {Bakkers},\ and\ \citenamefont {Kouwenhoven}}]{Mourik_2012}%
  \BibitemOpen
  \bibfield  {author} {\bibinfo {author} {\bibfnamefont {V.}~\bibnamefont
  {Mourik}}, \bibinfo {author} {\bibfnamefont {K.}~\bibnamefont {Zuo}},
  \bibinfo {author} {\bibfnamefont {S.~M.}\ \bibnamefont {Frolov}}, \bibinfo
  {author} {\bibfnamefont {S.~R.}\ \bibnamefont {Plissard}}, \bibinfo {author}
  {\bibfnamefont {E.~P. A.~M.}\ \bibnamefont {Bakkers}},\ and\ \bibinfo
  {author} {\bibfnamefont {L.~P.}\ \bibnamefont {Kouwenhoven}},\ }\href
  {https://doi.org/10.1126/science.1222360} {\bibfield  {journal} {\bibinfo
  {journal} {Science}\ }\textbf {\bibinfo {volume} {336}},\ \bibinfo {pages}
  {1003} (\bibinfo {year} {2012})}\BibitemShut {NoStop}%
\bibitem [{\citenamefont {Wang}\ \emph {et~al.}(2015)\citenamefont {Wang},
  \citenamefont {Zhou}, \citenamefont {Lian},\ and\ \citenamefont
  {Zhang}}]{Wang_2015}%
  \BibitemOpen
  \bibfield  {author} {\bibinfo {author} {\bibfnamefont {J.}~\bibnamefont
  {Wang}}, \bibinfo {author} {\bibfnamefont {Q.}~\bibnamefont {Zhou}}, \bibinfo
  {author} {\bibfnamefont {B.}~\bibnamefont {Lian}},\ and\ \bibinfo {author}
  {\bibfnamefont {S.-C.}\ \bibnamefont {Zhang}},\ }\href
  {https://doi.org/10.1103/PhysRevB.92.064520} {\bibfield  {journal} {\bibinfo
  {journal} {Phys. Rev. B}\ }\textbf {\bibinfo {volume} {92}},\ \bibinfo
  {pages} {064520} (\bibinfo {year} {2015})}\BibitemShut {NoStop}%
\bibitem [{\citenamefont {Lutchyn}\ \emph {et~al.}(2018)\citenamefont
  {Lutchyn}, \citenamefont {Bakkers}, \citenamefont {Kouwenhoven},
  \citenamefont {Krogstrup}, \citenamefont {Marcus},\ and\ \citenamefont
  {Oreg}}]{Lutchyn2018}%
  \BibitemOpen
  \bibfield  {author} {\bibinfo {author} {\bibfnamefont {R.~M.}\ \bibnamefont
  {Lutchyn}}, \bibinfo {author} {\bibfnamefont {E.~P. A.~M.}\ \bibnamefont
  {Bakkers}}, \bibinfo {author} {\bibfnamefont {L.~P.}\ \bibnamefont
  {Kouwenhoven}}, \bibinfo {author} {\bibfnamefont {P.}~\bibnamefont
  {Krogstrup}}, \bibinfo {author} {\bibfnamefont {C.~M.}\ \bibnamefont
  {Marcus}},\ and\ \bibinfo {author} {\bibfnamefont {Y.}~\bibnamefont {Oreg}},\
  }\href {https://doi.org/10.1038/s41578-018-0003-1} {\bibfield  {journal}
  {\bibinfo  {journal} {Nat. Rev. Mater.}\ }\textbf {\bibinfo {volume} {3}},\
  \bibinfo {pages} {52} (\bibinfo {year} {2018})}\BibitemShut {NoStop}%
\bibitem [{\citenamefont {Hasan}\ and\ \citenamefont
  {Kane}(2010)}]{Hasan&Kane_2010}%
  \BibitemOpen
  \bibfield  {author} {\bibinfo {author} {\bibfnamefont {M.~Z.}\ \bibnamefont
  {Hasan}}\ and\ \bibinfo {author} {\bibfnamefont {C.~L.}\ \bibnamefont
  {Kane}},\ }\href {https://doi.org/10.1103/RevModPhys.82.3045} {\bibfield
  {journal} {\bibinfo  {journal} {Rev. Mod. Phys.}\ }\textbf {\bibinfo {volume}
  {82}},\ \bibinfo {pages} {3045} (\bibinfo {year} {2010})}\BibitemShut
  {NoStop}%
\bibitem [{\citenamefont {Fu}\ \emph {et~al.}(2007)\citenamefont {Fu},
  \citenamefont {Kane},\ and\ \citenamefont {Mele}}]{Fu&Kane_2007}%
  \BibitemOpen
  \bibfield  {author} {\bibinfo {author} {\bibfnamefont {L.}~\bibnamefont
  {Fu}}, \bibinfo {author} {\bibfnamefont {C.~L.}\ \bibnamefont {Kane}},\ and\
  \bibinfo {author} {\bibfnamefont {E.~J.}\ \bibnamefont {Mele}},\ }\href@noop
  {} {\bibfield  {journal} {\bibinfo  {journal} {Phys. Rev. Lett.}\ }\textbf
  {\bibinfo {volume} {98}},\ \bibinfo {pages} {106803} (\bibinfo {year}
  {2007})}\BibitemShut {NoStop}%
\bibitem [{\citenamefont {Moore}\ and\ \citenamefont
  {Balents}(2007)}]{moore2007topological}%
  \BibitemOpen
  \bibfield  {author} {\bibinfo {author} {\bibfnamefont {J.~E.}\ \bibnamefont
  {Moore}}\ and\ \bibinfo {author} {\bibfnamefont {L.}~\bibnamefont
  {Balents}},\ }\href@noop {} {\bibfield  {journal} {\bibinfo  {journal} {Phys.
  Rev. B}\ }\textbf {\bibinfo {volume} {75}},\ \bibinfo {pages} {121306}
  (\bibinfo {year} {2007})}\BibitemShut {NoStop}%
\bibitem [{\citenamefont {Hsieh}\ \emph {et~al.}(2008)\citenamefont {Hsieh},
  \citenamefont {Qian}, \citenamefont {Wray}, \citenamefont {Xia},
  \citenamefont {Hor}, \citenamefont {Cava},\ and\ \citenamefont
  {Hasan}}]{Hsieh2008topological}%
  \BibitemOpen
  \bibfield  {author} {\bibinfo {author} {\bibfnamefont {D.}~\bibnamefont
  {Hsieh}}, \bibinfo {author} {\bibfnamefont {D.}~\bibnamefont {Qian}},
  \bibinfo {author} {\bibfnamefont {L.}~\bibnamefont {Wray}}, \bibinfo {author}
  {\bibfnamefont {Y.}~\bibnamefont {Xia}}, \bibinfo {author} {\bibfnamefont
  {Y.~S.}\ \bibnamefont {Hor}}, \bibinfo {author} {\bibfnamefont {R.~J.}\
  \bibnamefont {Cava}},\ and\ \bibinfo {author} {\bibfnamefont {M.~Z.}\
  \bibnamefont {Hasan}},\ }\href@noop {} {\bibfield  {journal} {\bibinfo
  {journal} {Nature}\ }\textbf {\bibinfo {volume} {452}},\ \bibinfo {pages}
  {970} (\bibinfo {year} {2008})}\BibitemShut {NoStop}%
\bibitem [{\citenamefont {Chen}\ \emph {et~al.}(2009)\citenamefont {Chen},
  \citenamefont {Analytis}, \citenamefont {Chu}, \citenamefont {Liu},
  \citenamefont {Mo}, \citenamefont {Qi}, \citenamefont {Zhang}, \citenamefont
  {Lu}, \citenamefont {Dai}, \citenamefont {Fang} \emph
  {et~al.}}]{3D_TI_experimental_2009}%
  \BibitemOpen
  \bibfield  {author} {\bibinfo {author} {\bibfnamefont {Y.}~\bibnamefont
  {Chen}}, \bibinfo {author} {\bibfnamefont {J.~G.}\ \bibnamefont {Analytis}},
  \bibinfo {author} {\bibfnamefont {J.-H.}\ \bibnamefont {Chu}}, \bibinfo
  {author} {\bibfnamefont {Z.}~\bibnamefont {Liu}}, \bibinfo {author}
  {\bibfnamefont {S.-K.}\ \bibnamefont {Mo}}, \bibinfo {author} {\bibfnamefont
  {X.-L.}\ \bibnamefont {Qi}}, \bibinfo {author} {\bibfnamefont
  {H.}~\bibnamefont {Zhang}}, \bibinfo {author} {\bibfnamefont
  {D.}~\bibnamefont {Lu}}, \bibinfo {author} {\bibfnamefont {X.}~\bibnamefont
  {Dai}}, \bibinfo {author} {\bibfnamefont {Z.}~\bibnamefont {Fang}}, \emph
  {et~al.},\ }\href@noop {} {\bibfield  {journal} {\bibinfo  {journal}
  {Science}\ }\textbf {\bibinfo {volume} {325}},\ \bibinfo {pages} {178}
  (\bibinfo {year} {2009})}\BibitemShut {NoStop}%
\bibitem [{\citenamefont {Liu}\ \emph {et~al.}(2010)\citenamefont {Liu},
  \citenamefont {Qi}, \citenamefont {Zhang}, \citenamefont {Dai}, \citenamefont
  {Fang},\ and\ \citenamefont {Zhang}}]{Modeling_2010}%
  \BibitemOpen
  \bibfield  {author} {\bibinfo {author} {\bibfnamefont {C.-X.}\ \bibnamefont
  {Liu}}, \bibinfo {author} {\bibfnamefont {X.-L.}\ \bibnamefont {Qi}},
  \bibinfo {author} {\bibfnamefont {H.}~\bibnamefont {Zhang}}, \bibinfo
  {author} {\bibfnamefont {X.}~\bibnamefont {Dai}}, \bibinfo {author}
  {\bibfnamefont {Z.}~\bibnamefont {Fang}},\ and\ \bibinfo {author}
  {\bibfnamefont {S.-C.}\ \bibnamefont {Zhang}},\ }\href
  {https://doi.org/10.1103/PhysRevB.82.045122} {\bibfield  {journal} {\bibinfo
  {journal} {Phys. Rev. B}\ }\textbf {\bibinfo {volume} {82}},\ \bibinfo
  {pages} {045122} (\bibinfo {year} {2010})}\BibitemShut {NoStop}%
\bibitem [{\citenamefont {de~Gennes}(1964)}]{DeGennes1964}%
  \BibitemOpen
  \bibfield  {author} {\bibinfo {author} {\bibfnamefont {P.~G.}\ \bibnamefont
  {de~Gennes}},\ }\href {https://doi.org/10.1103/RevModPhys.36.225} {\bibfield
  {journal} {\bibinfo  {journal} {Rev. Mod. Phys.}\ }\textbf {\bibinfo {volume}
  {36}},\ \bibinfo {pages} {225} (\bibinfo {year} {1964})}\BibitemShut
  {NoStop}%
\bibitem [{\citenamefont {Chung}\ \emph {et~al.}(2011)\citenamefont {Chung},
  \citenamefont {Qi}, \citenamefont {Maciejko},\ and\ \citenamefont
  {Zhang}}]{Chung_2011}%
  \BibitemOpen
  \bibfield  {author} {\bibinfo {author} {\bibfnamefont {S.~B.}\ \bibnamefont
  {Chung}}, \bibinfo {author} {\bibfnamefont {X.-L.}\ \bibnamefont {Qi}},
  \bibinfo {author} {\bibfnamefont {J.}~\bibnamefont {Maciejko}},\ and\
  \bibinfo {author} {\bibfnamefont {S.-C.}\ \bibnamefont {Zhang}},\ }\href
  {https://doi.org/10.1103/PhysRevB.83.100512} {\bibfield  {journal} {\bibinfo
  {journal} {Phys. Rev. B}\ }\textbf {\bibinfo {volume} {83}},\ \bibinfo
  {pages} {100512} (\bibinfo {year} {2011})}\BibitemShut {NoStop}%
\bibitem [{\citenamefont {He}\ \emph {et~al.}(2017)\citenamefont {He},
  \citenamefont {Pan}, \citenamefont {Stern}, \citenamefont {Burks},
  \citenamefont {Che}, \citenamefont {Yin}, \citenamefont {Wang}, \citenamefont
  {Lian}, \citenamefont {Zhou}, \citenamefont {Choi} \emph
  {et~al.}}]{He2017chiral}%
  \BibitemOpen
  \bibfield  {author} {\bibinfo {author} {\bibfnamefont {Q.~L.}\ \bibnamefont
  {He}}, \bibinfo {author} {\bibfnamefont {L.}~\bibnamefont {Pan}}, \bibinfo
  {author} {\bibfnamefont {A.~L.}\ \bibnamefont {Stern}}, \bibinfo {author}
  {\bibfnamefont {E.~C.}\ \bibnamefont {Burks}}, \bibinfo {author}
  {\bibfnamefont {X.}~\bibnamefont {Che}}, \bibinfo {author} {\bibfnamefont
  {G.}~\bibnamefont {Yin}}, \bibinfo {author} {\bibfnamefont {J.}~\bibnamefont
  {Wang}}, \bibinfo {author} {\bibfnamefont {B.}~\bibnamefont {Lian}}, \bibinfo
  {author} {\bibfnamefont {Q.}~\bibnamefont {Zhou}}, \bibinfo {author}
  {\bibfnamefont {E.~S.}\ \bibnamefont {Choi}}, \emph {et~al.},\ }\href@noop {}
  {\bibfield  {journal} {\bibinfo  {journal} {Science}\ }\textbf {\bibinfo
  {volume} {357}},\ \bibinfo {pages} {294} (\bibinfo {year}
  {2017})}\BibitemShut {NoStop}%
\bibitem [{\citenamefont {Kayyalha}\ \emph {et~al.}(2020)\citenamefont
  {Kayyalha}, \citenamefont {Xiao}, \citenamefont {Zhang}, \citenamefont
  {Shin},\ and\ \citenamefont {Chang}}]{2020Absence}%
  \BibitemOpen
  \bibfield  {author} {\bibinfo {author} {\bibfnamefont {M.}~\bibnamefont
  {Kayyalha}}, \bibinfo {author} {\bibfnamefont {D.}~\bibnamefont {Xiao}},
  \bibinfo {author} {\bibfnamefont {R.}~\bibnamefont {Zhang}}, \bibinfo
  {author} {\bibfnamefont {J.}~\bibnamefont {Shin}},\ and\ \bibinfo {author}
  {\bibfnamefont {C.~Z.}\ \bibnamefont {Chang}},\ }\href@noop {} {\bibfield
  {journal} {\bibinfo  {journal} {Science}\ }\textbf {\bibinfo {volume}
  {367}},\ \bibinfo {pages} {64} (\bibinfo {year} {2020})}\BibitemShut
  {NoStop}%
\bibitem [{\citenamefont {Nakajima}(1955)}]{S.Nakajima_1955}%
  \BibitemOpen
  \bibfield  {author} {\bibinfo {author} {\bibfnamefont {S.}~\bibnamefont
  {Nakajima}},\ }\href {https://doi.org/10.1080/00018735500101254} {\bibfield
  {journal} {\bibinfo  {journal} {Advances in Physics}\ }\textbf {\bibinfo
  {volume} {4}},\ \bibinfo {pages} {363} (\bibinfo {year} {1955})}\BibitemShut
  {NoStop}%
\bibitem [{\citenamefont {Schrieffer}\ and\ \citenamefont
  {Wolff}(1966)}]{Schrieffer&Wolff_1966}%
  \BibitemOpen
  \bibfield  {author} {\bibinfo {author} {\bibfnamefont {J.~R.}\ \bibnamefont
  {Schrieffer}}\ and\ \bibinfo {author} {\bibfnamefont {P.~A.}\ \bibnamefont
  {Wolff}},\ }\href {https://doi.org/10.1103/PhysRev.149.491} {\bibfield
  {journal} {\bibinfo  {journal} {Phys. Rev.}\ }\textbf {\bibinfo {volume}
  {149}},\ \bibinfo {pages} {491} (\bibinfo {year} {1966})}\BibitemShut
  {NoStop}%
\bibitem [{\citenamefont {Fr\"ohlich}(1950)}]{Frohlich_1950}%
  \BibitemOpen
  \bibfield  {author} {\bibinfo {author} {\bibfnamefont {H.}~\bibnamefont
  {Fr\"ohlich}},\ }\href {https://doi.org/10.1103/PhysRev.79.845} {\bibfield
  {journal} {\bibinfo  {journal} {Phys. Rev.}\ }\textbf {\bibinfo {volume}
  {79}},\ \bibinfo {pages} {845} (\bibinfo {year} {1950})}\BibitemShut
  {NoStop}%
\bibitem [{\citenamefont {Qiao}\ \emph {et~al.}(2022)\citenamefont {Qiao},
  \citenamefont {Li},\ and\ \citenamefont {Sun}}]{Qiao_2022}%
  \BibitemOpen
  \bibfield  {author} {\bibinfo {author} {\bibfnamefont {G.-J.}\ \bibnamefont
  {Qiao}}, \bibinfo {author} {\bibfnamefont {S.-W.}\ \bibnamefont {Li}},\ and\
  \bibinfo {author} {\bibfnamefont {C.~P.}\ \bibnamefont {Sun}},\ }\href
  {https://doi.org/10.1103/PhysRevB.106.104517} {\bibfield  {journal} {\bibinfo
   {journal} {Phys. Rev. B}\ }\textbf {\bibinfo {volume} {106}},\ \bibinfo
  {pages} {104517} (\bibinfo {year} {2022})}\BibitemShut {NoStop}%
\bibitem [{\citenamefont {Xia}\ \emph {et~al.}(2009)\citenamefont {Xia},
  \citenamefont {Qian}, \citenamefont {Hsieh}, \citenamefont {Wray},
  \citenamefont {Pal}, \citenamefont {Lin}, \citenamefont {Bansil},
  \citenamefont {Grauer}, \citenamefont {Hor}, \citenamefont {Cava} \emph
  {et~al.}}]{SurfaceObservation_2009}%
  \BibitemOpen
  \bibfield  {author} {\bibinfo {author} {\bibfnamefont {Y.}~\bibnamefont
  {Xia}}, \bibinfo {author} {\bibfnamefont {D.}~\bibnamefont {Qian}}, \bibinfo
  {author} {\bibfnamefont {D.}~\bibnamefont {Hsieh}}, \bibinfo {author}
  {\bibfnamefont {L.}~\bibnamefont {Wray}}, \bibinfo {author} {\bibfnamefont
  {A.}~\bibnamefont {Pal}}, \bibinfo {author} {\bibfnamefont {H.}~\bibnamefont
  {Lin}}, \bibinfo {author} {\bibfnamefont {A.}~\bibnamefont {Bansil}},
  \bibinfo {author} {\bibfnamefont {D.}~\bibnamefont {Grauer}}, \bibinfo
  {author} {\bibfnamefont {Y.~S.}\ \bibnamefont {Hor}}, \bibinfo {author}
  {\bibfnamefont {R.~J.}\ \bibnamefont {Cava}}, \emph {et~al.},\ }\href@noop {}
  {\bibfield  {journal} {\bibinfo  {journal} {Nature Physics}\ }\textbf
  {\bibinfo {volume} {5}},\ \bibinfo {pages} {398} (\bibinfo {year}
  {2009})}\BibitemShut {NoStop}%
\bibitem [{\citenamefont {Moore}\ \emph {et~al.}(2008)\citenamefont {Moore},
  \citenamefont {Ran},\ and\ \citenamefont {Wen}}]{WenXiaoGang_2008}%
  \BibitemOpen
  \bibfield  {author} {\bibinfo {author} {\bibfnamefont {J.~E.}\ \bibnamefont
  {Moore}}, \bibinfo {author} {\bibfnamefont {Y.}~\bibnamefont {Ran}},\ and\
  \bibinfo {author} {\bibfnamefont {X.-G.}\ \bibnamefont {Wen}},\ }\href
  {https://doi.org/10.1103/PhysRevLett.101.186805} {\bibfield  {journal}
  {\bibinfo  {journal} {Phys. Rev. Lett.}\ }\textbf {\bibinfo {volume} {101}},\
  \bibinfo {pages} {186805} (\bibinfo {year} {2008})}\BibitemShut {NoStop}%
\bibitem [{\citenamefont {Yu}\ \emph {et~al.}(2010)\citenamefont {Yu},
  \citenamefont {Zhang}, \citenamefont {Zhang}, \citenamefont {Zhang},
  \citenamefont {Dai},\ and\ \citenamefont {Fang}}]{RuiYu_2010}%
  \BibitemOpen
  \bibfield  {author} {\bibinfo {author} {\bibfnamefont {R.}~\bibnamefont
  {Yu}}, \bibinfo {author} {\bibfnamefont {W.}~\bibnamefont {Zhang}}, \bibinfo
  {author} {\bibfnamefont {H.-J.}\ \bibnamefont {Zhang}}, \bibinfo {author}
  {\bibfnamefont {S.-C.}\ \bibnamefont {Zhang}}, \bibinfo {author}
  {\bibfnamefont {X.}~\bibnamefont {Dai}},\ and\ \bibinfo {author}
  {\bibfnamefont {Z.}~\bibnamefont {Fang}},\ }\href
  {https://doi.org/10.1126/science.1187485} {\bibfield  {journal} {\bibinfo
  {journal} {Science}\ }\textbf {\bibinfo {volume} {329}},\ \bibinfo {pages}
  {61} (\bibinfo {year} {2010})}\BibitemShut {NoStop}%
\bibitem [{\citenamefont {Xu}\ \emph {et~al.}(2015)\citenamefont {Xu},
  \citenamefont {Wang}, \citenamefont {Liu}, \citenamefont {Ge}, \citenamefont
  {Yang}, \citenamefont {Liu}, \citenamefont {Xu}, \citenamefont {Guan},
  \citenamefont {Gao}, \citenamefont {Qian}, \citenamefont {Liu}, \citenamefont
  {Wang}, \citenamefont {Zhang}, \citenamefont {Xue},\ and\ \citenamefont
  {Jia}}]{XueQiKun2015}%
  \BibitemOpen
  \bibfield  {author} {\bibinfo {author} {\bibfnamefont {J.-P.}\ \bibnamefont
  {Xu}}, \bibinfo {author} {\bibfnamefont {M.-X.}\ \bibnamefont {Wang}},
  \bibinfo {author} {\bibfnamefont {Z.~L.}\ \bibnamefont {Liu}}, \bibinfo
  {author} {\bibfnamefont {J.-F.}\ \bibnamefont {Ge}}, \bibinfo {author}
  {\bibfnamefont {X.}~\bibnamefont {Yang}}, \bibinfo {author} {\bibfnamefont
  {C.}~\bibnamefont {Liu}}, \bibinfo {author} {\bibfnamefont {Z.~A.}\
  \bibnamefont {Xu}}, \bibinfo {author} {\bibfnamefont {D.}~\bibnamefont
  {Guan}}, \bibinfo {author} {\bibfnamefont {C.~L.}\ \bibnamefont {Gao}},
  \bibinfo {author} {\bibfnamefont {D.}~\bibnamefont {Qian}}, \bibinfo {author}
  {\bibfnamefont {Y.}~\bibnamefont {Liu}}, \bibinfo {author} {\bibfnamefont
  {Q.-H.}\ \bibnamefont {Wang}}, \bibinfo {author} {\bibfnamefont {F.-C.}\
  \bibnamefont {Zhang}}, \bibinfo {author} {\bibfnamefont {Q.-K.}\ \bibnamefont
  {Xue}},\ and\ \bibinfo {author} {\bibfnamefont {J.-F.}\ \bibnamefont {Jia}},\
  }\href {https://doi.org/10.1103/PhysRevLett.114.017001} {\bibfield  {journal}
  {\bibinfo  {journal} {Phys. Rev. Lett.}\ }\textbf {\bibinfo {volume} {114}},\
  \bibinfo {pages} {017001} (\bibinfo {year} {2015})}\BibitemShut {NoStop}%
\bibitem [{\citenamefont {Clayman}\ and\ \citenamefont
  {Frindt}(1971)}]{clayman1971superconducting}%
  \BibitemOpen
  \bibfield  {author} {\bibinfo {author} {\bibfnamefont {B.}~\bibnamefont
  {Clayman}}\ and\ \bibinfo {author} {\bibfnamefont {R.}~\bibnamefont
  {Frindt}},\ }\href@noop {} {\bibfield  {journal} {\bibinfo  {journal} {Solid
  State Communications}\ }\textbf {\bibinfo {volume} {9}},\ \bibinfo {pages}
  {1881} (\bibinfo {year} {1971})}\BibitemShut {NoStop}%
\bibitem [{\citenamefont {Chen}\ \emph {et~al.}(2010)\citenamefont {Chen},
  \citenamefont {Chu}, \citenamefont {Analytis}, \citenamefont {Liu},
  \citenamefont {Igarashi}, \citenamefont {Kuo}, \citenamefont {Qi},
  \citenamefont {Mo}, \citenamefont {Moore}, \citenamefont {Lu}, \citenamefont
  {Hashimoto}, \citenamefont {Sasagawa}, \citenamefont {Zhang}, \citenamefont
  {Fisher}, \citenamefont {Hussain},\ and\ \citenamefont {Shen}}]{2010massive}%
  \BibitemOpen
  \bibfield  {author} {\bibinfo {author} {\bibfnamefont {Y.~L.}\ \bibnamefont
  {Chen}}, \bibinfo {author} {\bibfnamefont {J.-H.}\ \bibnamefont {Chu}},
  \bibinfo {author} {\bibfnamefont {J.~G.}\ \bibnamefont {Analytis}}, \bibinfo
  {author} {\bibfnamefont {Z.~K.}\ \bibnamefont {Liu}}, \bibinfo {author}
  {\bibfnamefont {K.}~\bibnamefont {Igarashi}}, \bibinfo {author}
  {\bibfnamefont {H.-H.}\ \bibnamefont {Kuo}}, \bibinfo {author} {\bibfnamefont
  {X.~L.}\ \bibnamefont {Qi}}, \bibinfo {author} {\bibfnamefont {S.~K.}\
  \bibnamefont {Mo}}, \bibinfo {author} {\bibfnamefont {R.~G.}\ \bibnamefont
  {Moore}}, \bibinfo {author} {\bibfnamefont {D.~H.}\ \bibnamefont {Lu}},
  \bibinfo {author} {\bibfnamefont {M.}~\bibnamefont {Hashimoto}}, \bibinfo
  {author} {\bibfnamefont {T.}~\bibnamefont {Sasagawa}}, \bibinfo {author}
  {\bibfnamefont {S.~C.}\ \bibnamefont {Zhang}}, \bibinfo {author}
  {\bibfnamefont {I.~R.}\ \bibnamefont {Fisher}}, \bibinfo {author}
  {\bibfnamefont {Z.}~\bibnamefont {Hussain}},\ and\ \bibinfo {author}
  {\bibfnamefont {Z.~X.}\ \bibnamefont {Shen}},\ }\href@noop {} {\bibfield
  {journal} {\bibinfo  {journal} {Science}\ }\textbf {\bibinfo {volume}
  {329}},\ \bibinfo {pages} {659} (\bibinfo {year} {2010})}\BibitemShut
  {NoStop}%
\end{thebibliography}%

\end{document}